\documentclass[10pt, conference]{IEEEtran}

\usepackage{booktabs}
\usepackage{multirow}
\usepackage{tcolorbox}
\usepackage{balance}
\usepackage{colortbl}
\usepackage[shortlabels]{enumitem}
\usepackage{amsmath}
\usepackage{amsthm}
\usepackage{amsfonts}
\usepackage[square,numbers]{natbib}
\usepackage{url}
\usepackage{bm, bbm, dsfont}
\usepackage{adjustbox}
\usepackage{tabularx}
\usepackage{tikz}
\usepackage{subcaption}

\newcolumntype{L}{>{\raggedright\arraybackslash}X}

\definecolor{fill}{rgb}{0.85, 0.85, 0.85}

\newcommand{\barc}[1]{\begin{adjustbox}{width=.2\linewidth, height=.17cm}\begin{picture}(110,5) \linethickness{3pt}\put(10,2){\line(1, 0){#1}} \end{picture}\end{adjustbox}
}

\newcommand\ie{\textit{i.e.}}
\newcommand\eg{\textit{e.g.}}

\newcolumntype{L}{>{\raggedright\arraybackslash}X}

\def\BibTeX{{\rm B\kern-.05em{\sc i\kern-.025em b}\kern-.08em}}

\newtcolorbox{blockquote}{colback=blue!10,boxrule=0.5pt,colframe=black}

\begin{document}

\title{Lessons learned from hyper-parameter tuning for microservice candidate identification}

\author{
Rahul Yedida\textsuperscript{1}, Rahul Krishna\textsuperscript{2}, Anup Kalia\textsuperscript{2}, Tim Menzies\textsuperscript{1}, Jin Xiao\textsuperscript{2}, Maja Vukovic\textsuperscript{2}\\
\textsuperscript{1}Department of Computer Science, NC State, USA; 
\textsuperscript{2}IBM Research, USA\\
ryedida@ncsu.edu, \{rkrsn, anup.kalia\}@ibm.com, timm@ieee.org, \{jinoaix,maja\}@us.ibm.com}

\maketitle

\begin{abstract}
When optimizing software for the cloud, monolithic applications need to be partitioned into many
smaller \textit{microservices}. While many tools have been proposed for this task, we warn that the evaluation of those approaches 
has been incomplete; e.g. minimal prior exploration of  hyperparameter optimization.
Using a set of open source Java EE applications,
we show here that (a)~such optimization can significantly improve microservice partitioning; and that
(b)~an open issue for future work is  how to find which optimizer works best for different problems. To facilitate
that future work, 
see \url{https://github.com/yrahul3910/ase-tuned-mono2micro} for a reproduction package
for this research. 
\end{abstract}


\section{Introduction}
\pagestyle{plain}

Traditional software is
``monolithic''; i.e. one large entity that handles all concerns.
Such software needs to be divided into ``microservices''
to take full advantage of the flexibility and
scalability offered by cloud computing environments~\cite{larrucea2018microservices}. That microservice architecture allows for independent scalability, the use of different programming languages for each microservice, and does not have a single point of failure.


In our experience, the process of manually refactoring a monolithic application to a microservice architecture can be a time-consuming, expensive, and an error-prone endeavor. Different architects may propose different candidate sets when manually constructing a set of microservices.
Also, different software systems may have different inherent architectures.
Lastly, different engineers may have differing opinions on the number of microservices to split the software into. 
Hence, like many other researchers \cite{jin2019service,desai2021graph,kalia2020mono2micro,mazlami2017extraction} we explore publicly available machine learning methods and algorithms to identify a candidate set of microservices from a monolithic application. But even here, there are still problems. For example, as shown in column one of Table~\ref{tab:parameters}, there are multiple methods to find these partitions. Also, as seen in column two of that table, these methods contain one or more parameters that must be set via engineering judgment. However, different software architects (and different types of applications) may have different views on the best values for these configurations.


\begin{table}[t]
    \centering
    \caption{Hyper-parameter choices  studied in this paper.
    All the methods take runtime traces
(or uses cases) as their input}
    \label{tab:parameters} { \scriptsize
    \begin{tabularx}{\linewidth}{lLl}
        \toprule
        \textbf{Algorithm} & \textbf{Hyper-parameter} & \textbf{Range}  \\
        \midrule
        mono2micro \cite{kalia2020mono2micro} & Number of clusters & $\{2, 3, \ldots\}$  \\
        \midrule
        \multirow{5}{*}{FoSCI \cite{jin2019service}} & Number of clusters & $\{2, 3, \ldots\}$ \\
        & Number of iterations to run NSGA-II & $\{1, 2,\ldots\}$ \\
        & Population size for NSGA-II & $\{5, 6, \ldots\}$ \\
        & Parent size to use in NSGA-II & $\{5, 6, \ldots\}$ \\
        & Threshold for clustering to stop & $\mathbb{R}^+$ \\
        \midrule
        \multirow{2}{*}{MEM \cite{mazlami2017extraction}} & Maximum partition size & $\{2, 3,\ldots\}$ \\
        & Number of partitions & $\{2,3,\ldots\}$ \\
        \midrule
        \multirow{3}{*}{Bunch \cite{mitchell2006automatic}} & Number of partitions & $\{2,3,\ldots\}$ \\
        & Initial population size for hill climbing & $\{2, 3, \ldots\}$ \\
        & Number of neighbors to consider in hill-climbing iterations & $[0, 1]$ \\
        \midrule
        \multirow{4}{*}{CoGCN \cite{desai2021graph}} & Number of clusters & $\{2,3,\ldots\}$ \\
        & Loss function coefficients $\alpha_1, \alpha_2, \alpha_3$ & $\alpha_i \in \mathbb{R}$ \\
        & Number of hidden units in each layer, $h_1, h_2$ & $h_i \in \mathbb{Z}^+$ \\
        & Dropout rate & $d \in [0, 1)$ \\
        \bottomrule
    \end{tabularx} }
\end{table}

We characterize all these issues as {\em hyperparameter optimization} (HPO);
i.e. the automatic selection of (a)~which features/algorithms to use;
and (b)~what settings to apply to those algorithms.
In our case, the algorithms are the methods used to find partitions or microservices.
Using the partitioning methods in 
Table~\ref{tab:parameters}
the paper compares off-the-shelf partitioning 
methods to two tuning (\ie, HPO) methods. The rest of the paper does the following:
\begin{itemize}[wide=0pt]
\item Documents the improvement seen to microservice creation
when the methods of Table~\ref{tab:parameters} are tuned via
HPO. Specifically, tuning can significantly improve results, with the highest being 266\% (FoSCI on the plants dataset, on the SM metric).
\item
Reveals a new research challenge {\em microservice design strategies}.
While our results shows that tuning is better than non-tuning,
we also show that
no  single  approach  is  a  clear  winner  across  multiple datasets
or multiple performance goals.
\end{itemize}
In summary, the lessons learned from this work are:
\begin{enumerate}[leftmargin=*]
    \item \textbf{Do not use these partitioning
    methods   off-the-shelf.} since these
    these tend to have subpar results.
    \item \textbf{The choice of tuning algorithm is not as important as the decision to tune.} The random and hyperopt optimizers performed differently on different metrics and datasets, but in a broader view, shared the same number of highest Scott-Knott rankings. Therefore, the decision to tune is important, but the choice of algorithm may not be as important.
      \item \textbf{More research is required on tuning.}
      This work suggest that  a  new  research  challenge is appropriate:
{\em microservice design strategies}.  While  our  results  shows  that  tuning  is  better than  non-tuning,  we  also  show  that  no  single  approach is  a  clear  winner  across  multiple  datasets  or  multiple performance goals.
    \item \textbf{There is no best partitioning algorithm.} We will show later in this paper that the priorities of the stakeholders matters when choosing the right algorithm, and that there is no one best algorithm across all the metrics. 
\end{enumerate}
More generally, going forward, we say that industrial cloud applications need to understand and apply
  automatic tuning methods. That said, we will conclude that  it is still
an open issue which tuner to use. More research is needed in that direction.

\section{Business Case}
\label{sec:background}

A recent report shows that only 20\% of the enterprise workloads are in the cloud, and they were predominately written for native cloud architectures~\cite{daya2016microservices}. This leaves 80\% of legacy applications on-premises, waiting to be refactored  and modernized for the cloud. Once ported to microservices, business services can be independently enhanced and scaled, providing agility and improved speed of delivery.

{Application refactoring} is the process of restructuring existing code into a set of smaller independent groups of code, ideally without changing the external behavior and semantics. Currently, refactoring is usually done manually and is expensive, time-consuming, and error-prone.
There are several methods to implement
application refactoring. For example, Table~\ref{tab:parameters} lists different ways to partition
a monolith. In the table,  the core
idea is to understand what parts
 of the code call each other other
 (e.g., by reflecting on test cases
 or uses cases), then cluster those
 parts of the code into separate microservices.
 
Software \textit{remodularization} aims to repair the architectural design of a software system to reduce the technical debt. Remodularization tools such as Bunch \cite{mitchell2006automatic} perform such refactoring with the underlying assumption that developers want to maximize cohesion and minimize coupling \cite{candela2016using}.
  

As opposed to service-oriented architectures, the construction of microservices (as explored here) is motivated by enterprises' need to refactor their code into independent services, guided by some automatically-inferred partitions. The resulting application architecture may or may not conform perfectly to SOA, as the partitions are reversed engineered from test cases
or use cases describing taken from some pre-exisiting monolithic application. The benefit is these microservices can be created relatively cheaply
and built incrementally by porting existing
monolithic code to the cloud. Well observed practices in the industry, such as the strangler pattern, follows this approach.





\section{Background}

Several approaches have been proposed to solve this problem. These may broadly be categorized into (a) Optimization (or search) based approaches, and (b) unsupervised clustering based approaches. We provide a brief overview of each of these approaches below.

\subsection{Unsupervised Clustering Based Partitioning}
\subsubsection{Mono2Micro} This approach, introduced by~\citet{kalia2020mono2micro}, uses hierarchical decomposition that dynamically collects runtime traces, under the execution of specific business use cases of the application. Then, using Jaccard similarity and hierarchical clustering over the runtime traces, the applications classes are segregated into disjoint partitions.

\subsubsection{Microservice Extractor (aka. MEM)} \citet{mazlami2017extraction} propose MEM, which first constructs a graph based on logical, semantic, and change history based coupling. Next, they pose microservice partitioning as a graph cut problem that uses minimum spanning trees to find min-cut partitions from the constructed graph. Classes in each min-cut represents a candidate microservice partition.

\subsection{Optimization/Search Based Partitioning}

Some authors prefer to cast software partitioning as an optimization problem, whose objective is to optimize a linear combination of the quality metrics.
 Our reading of the literature is that the following list of
 methods  samples much of the current state-of-the-art in the research arena.

\subsubsection{CoGCN}~\citet{desai2021graph} proposed a method called CO-GCN, that uses graph convolutional networks (GCN) \cite{kipf2016semi} with a loss function that incorporates several metrics from Tab.~\ref{tab:metrics} directly the end-to-end training phase. The partitions are then discovered using the k-means clustering approach. 

\subsubsection{FoSCI}~\citet{jin2019service} propose FoSCI, which uses runtime traces as its data source. They first remove redundant traces using a trace reduction algorithm, and then perform hierarchical clustering to create ``functional atoms'' (logical units that provide some function). These functional atoms are then merged using a multi-objective optimization algorithm, namely NSGA-II~\cite{deb2002fast}.


\subsubsection{Bunch}~\citet{mitchell2006automatic} proposed Bunch, where the partition problem is characterized as a search problem. Computational search procedures are then used to find a partition that maximizes the modular quality (MQ) and the structural modular quality (SMQ) metrics. In particular, they use the hill climbing search algorithm~\cite[\S4.1]{mitchell2006automatic}.

\subsection{Hyper-parameter optimization}
We note that across all of these approaches, a common theme is the existence of hyper-parameters, listed in Table \ref{tab:parameters}. Because these papers are focused on proposing a novel approach, little attention is paid to tuning these hyper-parameters, or discussing reasonable defaults for them 
Therefore, we ask whether tuning can improve the results. 

Hyper-parameter optimizers find best suited parameters (\ie, meta-options) that guide the specifics of an algorithm's execution. These include parameters such as the number of clusters for a partitioning or a clustering algorithm, or the number of layers in a deep learner. 
 
\citet{bergstra2012random} advocate for random hyper-parameter tuning, in part, due to the ease of implementation.  
A more sophisticated optimizer is the Tree of Parzen Estimators (TPE) \cite{bergstra2011algorithms} that uses  an ``acquisition function'' to reflect on the model built so far in order to learn the next most informative sample to be evaluated. Prior results are divided into {\em best} and {\em rest} and each division is modelled as a Gaussian distribution. Candidate evaluations are then ranked by how well they select for {\em best} and avoid {\em rest}. The better candidates are then evaluated which, in turn, updates {\em best} and {\em rest} and their associated distributions. We use the hyperopt \cite{komer2014hyperopt} implementation of the TPE algorithm, and in the rest of the paper, we use ``hyperopt" to refer to ``TPE".

\begin{table*}[!t]
    \centering
    \caption{The metrics used to assess the quality of microservice partitions. In the first column, a $\Downarrow$ means we want to minimize the metric; a $\Uparrow$ means that higher values are better.}
    \label{tab:metrics}
    \begin{tabularx}{\textwidth}{llLL}
        \toprule
        \textbf{Metric} & \textbf{Name} & \textbf{Description} & \textbf{Goal} \\
        \midrule
        BCP $\Downarrow$ \cite{kalia2020mono2micro} & Business context purity & Mean entropy of business use cases per partition & If \textbf{minimized} then more business cases handled locally \\
        ICP $\Downarrow$ \cite{kalia2020mono2micro} & Inter-partition call percentage & Percentage of runtime calls that are between partitions & If \textbf{minimized} then less traffic between clusters \\
        SM $\Uparrow$ \cite{jin2019service} & Structural modularity & A combination of cohesion and coupling defined by \citet{jin2019service} & If \textbf{maximized} then more self-contained clusters with fewer connections between them \\
        MQ $\Uparrow$ \cite{mitchell2006automatic} & Modular quality & A ratio involving cohesion and coupling, defined by \citet{mitchell2006automatic} & If \textbf{maximized} then more processing is local to a cluster \\
        IFN $\Downarrow$ \cite{mitchell2006automatic} & Interface number & Number of interfaces needed in the microservice architecture & If \textbf{minimized} then fewer calls between clusters \\ 
        \bottomrule
    \end{tabularx}
\end{table*}

Several papers~\cite{fu2016tuning,nair2018finding,agrawal2021simpler,yedida2021value} demonstrate
that the hyper-parameters learned via such optimizers can lead to models that significantly out-perform the untuned
(or off-the-shelf) defaults.  Yet, none of the approaches we study do such tuning. We investigate this research gap by using two of the optimizers recommended by \citet{bergstra2011algorithms}: random and TPE.  Both of these optimizers have been endorsed in major venues (NeurIPS) and have since been widely used by the community. 

\begin{table*}[t!]
    \caption{Summary of Results. Each column is a separate clustering method. Each row is a different performance metric. The final row highlights the wins across all metrics. }
    \label{tab:summary-only}
    \begin{adjustbox}{width=0.9\textwidth}
   { \scriptsize
    \begin{subtable}{0.12\textwidth}
        \centering
        \caption{mono2micro}
        \label{tab:mono2micro:summary}
        \begin{tabular}{lll}
            \toprule 
            \textbf{Metric} & \textbf{Treatment} & \textbf{Wins} \\
            \midrule
            \multirow{3}{*}{BCP $\Downarrow$} & untuned & 0 \\
            & random & 4 \barc{100} \\
            & hyperopt & 4 \barc{100} \\
            \midrule
            \multirow{3}{*}{ICP $\Downarrow$} & untuned & 4 \barc{100} \\
            & random & 0 \\
            & hyperopt & 0 \\
            \midrule
            \multirow{3}{*}{SM $\Uparrow$} & untuned & 2 \barc{50} \\
            & random & 3 \barc{75} \\
            & hyperopt & 3 \barc{75} \\
            \midrule 
            \multirow{3}{*}{MQ $\Uparrow$} & untuned & 1 \barc{25} \\
            & random & 3 \barc{75} \\
            & hyperopt & 3 \barc{75} \\
            \midrule
            \multirow{3}{*}{IFN $\Downarrow$} & untuned & 1 \barc{25} \\
            & random & 4 \barc{100} \\
            & hyperopt & 4 \barc{100} \\
            \midrule
            \multirow{3}{*}{\textbf{Total}} & untuned & 1 \barc{20} \\
            & \cellcolor{blue!12}tuned & \cellcolor{blue!12}4 \barc{80} \\
            & tie & 0 \\
            \bottomrule
        \end{tabular}
    \end{subtable} \hspace{0.09\textwidth}
    \begin{subtable}{0.12\textwidth}
        \centering
        \caption{FoSCI}
        \label{tab:fosci:summary}
        \begin{tabular}{lll}
            \toprule 
            \textbf{Metric} & \textbf{Treatment} & \textbf{Wins} \\
            \midrule
            \multirow{3}{*}{BCP $\Downarrow$} & untuned & 0 \\
            & random & 0 \\
            & hyperopt & 4 \barc{100} \\
            \midrule
            \multirow{3}{*}{ICP $\Downarrow$} & untuned & 1 \barc{25} \\
            & random & 3 \barc{75} \\
            & hyperopt & 0 \\
            \midrule
            \multirow{3}{*}{SM $\Uparrow$} & untuned & 0 \\
            & random & 4 \barc{100} \\
            & hyperopt & 0 \\
            \midrule 
            \multirow{3}{*}{MQ $\Uparrow$} & untuned & 0 \\
            & random & 0 \\
            & hyperopt & 4 \barc{100} \\
            \midrule
            \multirow{3}{*}{IFN $\Downarrow$} & untuned & 1 \barc{25} \\
            & random & 4 \barc{100} \\
            & hyperopt & 1 \barc{25} \\
            \midrule
            \multirow{3}{*}{\textbf{Total}} & untuned & 0 \\
            & \cellcolor{blue!12}tuned & \cellcolor{blue!12}5 \barc{100} \\
            & tie & 0 \\
            \bottomrule
        \end{tabular}
    \end{subtable} \hspace{0.09\textwidth}
    \begin{subtable}{0.12\textwidth}
        \centering
        \caption{MEM}
        \label{tab:mem:summary}
        \begin{tabular}{lll}
            \toprule 
            \textbf{Metric} & \textbf{Treatment} & \textbf{Wins} \\
            \midrule
            \multirow{3}{*}{BCP $\Downarrow$} & untuned & 3 \barc{75} \\
            & random & 0 \\
            & hyperopt & 1 \barc{25} \\
            \midrule
            \multirow{3}{*}{ICP $\Downarrow$} & untuned & 2 \barc{50} \\
            & random & 2 \barc{50} \\
            & hyperopt & 0 \\
            \midrule
            \multirow{3}{*}{SM $\Uparrow$} & untuned & 3 \barc{75} \\
            & random & 0 \\
            & hyperopt & 1 \barc{25} \\
            \midrule 
            \multirow{3}{*}{MQ $\Uparrow$} & untuned & 1 \barc{25} \\
            & random & 2 \barc{50} \\
            & hyperopt & 4 \barc{100} \\
            \midrule
            \multirow{3}{*}{IFN $\Downarrow$} & untuned & 2 \barc{50} \\
            & random & 2 \barc{50} \\
            & hyperopt & 1 \barc{25} \\
            \midrule
            \multirow{3}{*}{\textbf{Total}} & \cellcolor{blue!12}untuned & \cellcolor{blue!12}{2 \barc{40}} \\
            & tuned & 1 \barc{20} \\
            & tie & 2 \barc{40} \\
            \bottomrule
        \end{tabular}
    \end{subtable} \hspace{0.09\textwidth}
    \begin{subtable}{0.12\textwidth}
        \centering
        \caption{Bunch}
        \label{tab:bunch:summary}
        \begin{tabular}{lll}
            \toprule 
            \textbf{Metric} & \textbf{Treatment} & \textbf{Wins} \\
            \midrule
            \multirow{3}{*}{BCP $\Downarrow$} & untuned & 2 \barc{50} \\
            & random & 3 \barc{75} \\
            & hyperopt & 3 \barc{75} \\
            \midrule
            \multirow{3}{*}{ICP $\Downarrow$} & untuned & 2 \barc{50} \\
            & random & 4 \barc{100} \\
            & hyperopt & 4 \barc{100} \\
            \midrule
            \multirow{3}{*}{SM $\Uparrow$} & untuned & 4 \barc{100} \\
            & random & 2 \barc{50} \\
            & hyperopt & 2 \barc{50} \\
            \midrule 
            \multirow{3}{*}{MQ $\Uparrow$} & untuned & 3 \barc{75} \\
            & random & 3 \barc{75} \\
            & hyperopt & 3 \barc{75} \\
            \midrule
            \multirow{3}{*}{IFN $\Downarrow$} & untuned & 2 \barc{50} \\
            & random & 4 \barc{100} \\
            & hyperopt & 4 \barc{100} \\
            \midrule
            \multirow{3}{*}{\textbf{Total}} & untuned & 1 \barc{20} \\
            & \cellcolor{blue!12}tuned & \cellcolor{blue!12}3 \barc{60} \\
            & tie & 1 \barc{20} \\
            \bottomrule
        \end{tabular}
    \end{subtable} \hspace{0.09\textwidth}
    \begin{subtable}{0.12\textwidth}
        \centering
        \caption{CoGCN}
        \label{tab:cogcn:summary}
        \begin{tabular}{lll}
            \toprule 
            \textbf{Metric} & \textbf{Treatment} & \textbf{Wins} \\
            \midrule
            \multirow{3}{*}{BCP $\Downarrow$} & untuned & 2 \barc{50} \\
            & random & 0 \\
            & hyperopt & 2 \barc{50} \\
            \midrule
            \multirow{3}{*}{ICP $\Downarrow$} & untuned & 0 \\
            & random & 0 \\
            & hyperopt & 4 \barc{100} \\
            \midrule
            \multirow{3}{*}{SM $\Uparrow$} & untuned & 0 \\
            & random & 2 \barc{50} \\
            & hyperopt & 2 \barc{50} \\
            \midrule 
            \multirow{3}{*}{MQ $\Uparrow$} & untuned & 4 \barc{100} \\
            & random & 0 \\
            & hyperopt & 0 \\
            \midrule
            \multirow{3}{*}{IFN $\Downarrow$} & untuned & 1 \barc{25} \\
            & random & 0 \\
            & hyperopt & 3 \barc{75} \\
            \midrule
            \multirow{3}{*}{\textbf{Total}} & untuned & 1 \barc{20} \\
            & \cellcolor{blue!12}tuned & \cellcolor{blue!12}3 \barc{60} \\
            & tie & 1 \barc{20} \\
            \bottomrule
        \end{tabular}
    \end{subtable}
    } \end{adjustbox}
\end{table*}

\begin{table}[]
    \centering
    \caption{Another summary of the results. Here, we show the best algorithm for an dataset for a metric (i.e., the best of untuned, random, and hyperopt). Where untuned had the best result, we mark it with an asterisk. A \colorbox{blue!10}{blue} background indicates the best result. The last row denotes the number of wins for each algorithm across all projects.}
    \label{tab:summary-metrics}
    { \scriptsize
    \begin{tabular}{lllllll}
        \toprule
        \textbf{Dataset} & \textbf{Algorithm} & \textbf{BCP $\Downarrow$} & \textbf{ICP $\Downarrow$} & \textbf{SM $\Uparrow$} & \textbf{MQ $\Uparrow$} & \textbf{IFN $\Downarrow$} \\
        \midrule
        \multirow{5}{*}{jpetstore} & mono2micro & \cellcolor{blue!12}{1.67} & 0.51* & 0.06* & 3.29 & 3.77 \\
        & Bunch & 2.43 & 0.48 & 0.21 & \cellcolor{blue!12}{7.95} & 2.64 \\
        & CO-GCN & 1.97* & 0.15 & 0.19 & 6.11* & 1.8 \\
        & MEM & 2.69* & 0.47* & \cellcolor{blue!12}{0.23*} & 4.56 & 4 \\
        & FoSCI & 2.04 & \cellcolor{blue!12}{0.04} & 0.12 & 3.63 & \cellcolor{blue!12}{1} \\
        \midrule
        \multirow{5}{*}{plants} & mono2micro & \cellcolor{blue!12}{1.73} & 0.51* & 0.1 & 2.82* & 5.1 \\
        & Bunch & 1.75 & 0.51 & 0.22* & \cellcolor{blue!12}{4.71} & 3.88 \\
        & CO-GCN & 1.87 & \cellcolor{blue!12}{0.02} & \cellcolor{blue!12}{0.27} & 3.51* & \cellcolor{blue!12}{1} \\
        & MEM & 2.04 & 0.32 & 0.25 & 2.83 & 4.17 \\
        & FoSCI & 2.04 & 0.08 & 0.11 & 3.63 & 3 \\
        \midrule 
        \multirow{5}{*}{daytrader} & mono2micro & 1.4 & 0.4* & 0.07 & 5.52 & 6.13 \\
        & Bunch & 1.75 & 0.51 & \cellcolor{blue!12}{0.27*} & \cellcolor{blue!12}{11.33*} & 3.88 \\
        & CO-GCN & \cellcolor{blue!12}{1.02*} & \cellcolor{blue!12}{0.01} & 0.15 & 7.99* & \cellcolor{blue!12}{0.5} \\
        & MEM & 1.94* & 0.38 & 0.11 & 4.73 & 4.08 \\
        & FoSCI & 2.04 & 0.04 & 0.12 & 0.12 & 1 \\
        \midrule 
        \multirow{5}{*}{acmeair} & mono2micro & \cellcolor{blue!12}{1.04} & 0.4* & 0.1 & 3.44 & 3.3 \\
        & Bunch & 1.67* & 0.51 & \cellcolor{blue!12}{0.17} & \cellcolor{blue!12}{4.71} & 3.88 \\
        & CO-GCN & 1.31 & \cellcolor{blue!12}{0.29} & 0.15 & 3.71* & \cellcolor{blue!12}{3} \\
        & MEM & 1.92* & 0.66 & 0.09 & 2.77* & 4 \\
        \midrule
        \multirow{5}{*}{\textbf{Wins}} & mono2micro & \cellcolor{blue!12}{3} & 0 & 0 & 0 & 0 \\
        & Bunch & 0 & 0 & \cellcolor{blue!12}{2} & \cellcolor{blue!12}{4} & 0 \\
        & CO-GCN & 1 & \cellcolor{blue!12}{2} & 1 & 0 & \cellcolor{blue!12}{2} \\
        & MEM & 0 & 1 & 1 & 0 & 1 \\
        & FoSCI & 0 & \cellcolor{blue!12}{2} & 0 & 0 & \cellcolor{blue!12}{2} \\
        \bottomrule
    \end{tabular}
    }
\end{table}

\section{Experimental Setup}
\label{sec:method}

We run each algorithm using the default settings (untuned) and using the settings obtained from hyper-parameter optimizations (tuned). We run each algorithm for 30 times both for tuned and untuned settings.

\noindent{\bf Data.}~We use four open-source projects to test the algorithms under study. These are: (1) \textit{daytrader}\footnote{\url{https://github.com/WASdev/sample.daytrader7}}, a sample online stock trading system; (2)~\textit{plants}\footnote{\url{https://github.com/WASdev/sample.plantsbywebsphere}}, a sample online store to purchase plants and pets; (3)~\textit{acmeair}\footnote{\url{https://github.com/acmeair/acmeair}}, a sample airline booking application; and (4) \textit{jpetstore}\footnote{\url{https://github.com/mybatis/jpetstore-6}}, a sample pet store. These applications are representative web apps built using various Java EE frameworks such as Spring, Apache Struts, etc.

\noindent{\bf Metrics.}~For a fair comparison with prior work, we must compare using the same set of metrics. We choose a total of five metrics to evaluate our core hypothesis that hyper-parameter tuning improves microservice extraction algorithms. These are detailed in Table \ref{tab:metrics}. These metrics have been used in prior studies, although different papers used different metrics in their evaluations. For fairness, we use metrics from all prior papers (cf.~Table \ref{tab:metrics}). These metrics evaluate different aspects of the utility of an algorithm that might be more useful to different sets of users (detailed in the RQ1), \eg, BCP evaluates how well different business use cases are separated across the microservices.

\noindent{\bf Hyper-parameter optimization.}~We use two hyper-parameter optimization approaches, random sampling \cite{bergstra2012random}, and Tree of Parzen Estimators (TPE) \cite{bergstra2011algorithms}, using the hyperopt framework \cite{bergstra2013hyperopt}. These are two commonly used approaches from the AI literature, and have been compared against in recent SE literature \cite{agrawal2021simpler}. We run both for 100 iterations.
We use two hyper-parameter optimization approaches for the following reasons. Firstly, it avoids discussing the \textit{choice} of hyper-parameter optimizers by demonstrating the use of two different approaches. Our argument is for the use of hyper-parameter optimization in general, as opposed to a specific approach.

Because we use several metrics, we need to guide the hyper-parameter optimization algorithms to achieve the results that we want. To achieve this, we 
define a \textit{loss} function $\ell(P) = \sum\limits_{i=1}^5 \alpha_i m_i(P)$, where $P$ is the partition, $m_i$ are the five metrics. For simplicity, we set each weight to 1 if the goal is to \textit{minimize} the metric, and -1 if the goal is to \textit{maximize} the metric. Having framed our loss function, we use the hyper-parameter optimization algorithms to minimize this goal. We note that this is not necessarily the \textit{best} loss function; but it is the simplest.

\noindent{\bf Statistics.}~Our comparison method of choice for the results is the Scott-Knott test, which was endorsed at ICSE'15 \cite{ghotra2015revisiting}. The Scott-Knott test is a recursive bi-clustering algorithm that terminates when the difference between the two split groups is insignificant. Scott-Knott searches for split points that maximize the expected value of the difference between the means of the two resulting groups. The result of the Scott-Knott test is \textit{ranks} assigned to each result set
(two results have the same rank if their difference is insignificant). 

\section{Results}
\label{sec:results}
To evaluate the outcome of hyperparameter tuning for microservice candidate identification, we ask and answer the following thse research questions:
\begin{itemize}[wide=0pt]
\item {\bf RQ1.~}~\textit{How does hyperparameter tuning affect microservice candidate identification?}
\item {\bf RQ2.~}~\textit{Is there an exemplary approach that outperforms the others across multiple datasets?}
\end{itemize}
For brevity we show only the summary tables in this paper. We provide the details online\footnote{\url{https://github.com/yrahul3910/tuned-microservices-results}}.

\noindent \textbf{RQ1. How does hyperparameter tuning affect microservice candidate identification?}

Table~\ref{tab:summary-only} shows our results.  In that figure, each row shows results from one assessment metric (BCP, ICP, SM, MQ, IDM) and each column shows results from
a different technology from Table~1. Each result is a three bars chart showing how often the untuned, random tuned, and TPE (hyperopt) tuner achieved best results   first (where ``best'' was determined by our statistical tests)  within our four data sets (daytrader, plants, acmeair, jpetstore). For example, top-left, we see that random and
hyperopt both achieved best BCP scores  for mono2micro. 

The general trend across all the results is (a)~untuned usually lost to tuning; but (b) there is no stable pattern about which tuner (random or hyperopt) worked best for a particular performance measure.   The exception to this general pattern is that (c)~untuned MEM which performed best in 3/5 metrics (BCP, ICP, SM). Our investigation of MEM revealed that it ignores its hyperparameter options
and uses its own internal methods for deciding actual parameters. 

\begin{blockquote}
    \noindent \textbf{Summary.~}
   Tuning is better than not tuning, but it is still an open issue is a particular tuner works better. 
\end{blockquote}

\noindent \textbf{RQ2. Is there an exemplary approach that outperforms the others across multiple datasets?}

Our review of different approaches in \S \ref{sec:background} showed a variety of techniques applied to the problem. These tools were all available as open-source, and could be easily run and evaluated. Table \ref{tab:summary-metrics} shows a summary view of the results. Here, we pick the best among the untuned and tuned results (from RQ1) and compare the algorithms across all the datasets and metrics. We see that there is no one best algorithm, and that the ``best" algorithm depends on the specific use case and metric used to reflect that use case. For example, a business user might be more interested in the BCP metric (i.e., the business cases are more cleanly separated), and would be recommended the \textit{mono2micro} algorithm. Similarly, a systems engineer who wants low coupling (i.e., low ICP and low IFN) would choose the CoGCN or the FoSCI approach. Therefore, the conclusion is that the right algorithm depends on the user goals and priorities.
\begin{blockquote}
    \noindent \textbf{Summary.~}The best partitioning algorithm depends on the business use case of interest to the practitioners.
\end{blockquote}

{\footnotesize  \bibliographystyle{IEEEtranSN}
\balance
\bibliography{cite}
}

\end{document}